# A framework for designing concurrent and recoverable abstract data types based on commutativity

Carmelo MALTA, José MARTINEZ

Université des Sciences et Techniques du Languedoc
Centre de Recherche en Informatique de Montpellier
860, rue de Saint-Priest, 34090 Montpellier, FRANCE
Email : <malta, martinez>@crim.fr

**Abstract**
In this paper, we try to focus the reader's interest on the problems that transactional systems have to resolve for taking advantage of commutativity in a serializable and recoverable way. Our framework is, (as others), based on the use of conditional commutativity on abstract date types. We present new features that have not been found in the literature hitherto, that both increase concurrency and simplify recovery.

## 1. INTRODUCTION

One can view the methods of controlling the concurrent accesses to shared data as belonging to a large scale going from syntactic methods, (like strict two-phase locking [5] with read and write locks, which is the only correct method when no knowledge at all is provided on data), to semantic methods, (among which the most semantic of all is parallel programming, e. g., with CSP [9]).

The evolution is naturally to propose more and more semantic methods. The challenge is double: A trend is to increase the knowledge that the concurrency control has of the semantics of the data, then to exploit this knowledge in order to increase the parallelism within and between transactions.

In most cases, semantics is added at the level of individual objects. Propositions generally increase the set of locks on databases with granular organizations [7, 10], of operations on typed objects [11, 18], in object-oriented databases [6, 4, 14]. Although the system has a more precise view, the criterion used remains serializability because compatibility or commutativity is used at the object level. Badrinath & Ramamritham [2] introduce the relative recoverability criterion which still implies serializability at the transaction level.

---
This work was supported in part by the PRC-BD3 coordinated by the Institut National de la Recherche en Informatique et Automatique (INRIA), and in part by the PRC-C3 coordinated by the Centre National de la Recherche Scientifique (CNRS).

Other propositions introduce new criteria at the transaction level: multi-level atomicity [13] or "acceptability" of compensative transactions [8, 19]. These lead to non-serializable schedules.

In this paper, we discuss the specific case of commutativity of operations on abstract data types (ADTs). However, we do not just discuss concurrency control, but also recovery, in a serializable and cascading rollback-free way.

The paper is organized as follows: In section 2, we introduce the main point of our framework for dealing both with concurrency control and recovery: a two-level interface. In section 3, first we discuss commutativity itself. Next, we present our general protocol, at the object level, which uses a monitor rather than locking in order to allow conditional commutativity with out-parameters without executing more than once any operation. Finally, in the conclusion, we review the advantages and disadvantages of our scheme and introduce some luxury features added in our implementation.

## 2. A TWO-LEVEL INTERFACE FOR ADTs

All the features that we present in our framework are based on the requirements needed to carrying out both concurrency control and recovery. Our approach is characterized by:

(i) a unique copy of the object, which is modified concurrently and maintained up-to-date;

(ii) a recovery mechanism based on inverse operations;

(iii) two interfaces for the operations on an ADT: a first one, visible to the user of the ADT; a second one, hidden to the user, for dealing with concurrency control and recovery;

(iv) a concurrency control mechanism which only takes into account the in- and out-parameters of the operations but never (directly) the state of the object;

(v) the possibility to return from the invocation of an operation without executing it, (using deduced return values).

In the sequel, we present the two-level interface used for operations on an ADT, and the advantages which it contributes to for recovery and concurrency control.

### 2.1. Motivations
The reason why we need a two-level interface is essentially motivated by recovery. It is also influenced by the fact that we do not determine commutativity of operations by considering directly the *current state* of an object. This self-imposed constraint is relevant for big objects which cannot be analyzed *a priori*, but it deserves to be removed when objects are sufficiently small. It is the case with the Escrow method [15], which takes into account the initial value of an aggregate, (integer or real), to increase commutativity of increment and decrement operations. However, considering out-parameters of operations is generally sufficient to have an idea of the *initial state* of the object; Rœssler & Burkhard [16] named it the "peephole" approach. Besides, the state of an object is just an abbreviation of the sequence of applied and non-rejected operations.

We also take advantage of this two-level interface for simplifying concurrency control.

### 2.2. State-based inverse operations
As stated in (i), the object state is maintained up-to-date, i. e., every operation is performed on a single copy. As noted by Weihl [22], such a choice implies the use of inverse operations in order to treat recovery adequately. Now, what is to be clear is that inverse operations cannot be state-independent.

*Why state-based recovery?*
Let us take the generic SET example. We consider only the INSERT operation for the purpose of our demonstration. From the specification of a SET, it might be deduced that INSERT commutes with itself. This is false because a transaction which invokes an INSERT operation on a given SET may abort at a later time. Therefore, we must provide an inverse operation. This inverse is either DELETE, if the INSERT was successful, or NULL, if the inserted item was already in the SET.

It is worth considering that virtually there exists two kinds of INSERT operations: one defined from a SET which does not contain the inserted item, and a second defined from a SET which already contains the item. Each of these conceptually different operations has its own inverse, respectively DELETE and NULL. Neither two INSERTs of the first category can commute on a common item, nor an INSERT of the first category with one of the other, whereas it is allowed for two INSERTs belonging to the second category.

Trying to enhance concurrency by allowing commutativity between two INSERTs of the first category, (because the report is unknown to the outside), yields to context-sensitivity, i. e., an inverse operation is no more associated to just one direct operation but to a set of operations. In this example, when there is more than one INSERT of the first category, the inverse is NULL; for the last INSERT, however, the inverse is DELETE. We know of no protocol which implements such a complex mechanism.

*A private interface*
As the out-parameters given for the operations of an ADT may not be sufficient to determine the corresponding inverse operations, we introduce a second interface. This new interface is used by the transaction manager, and a mapping from the interface known by the user, the public one, to the private one is given. This point is developed in the next section.

It is also possible to explain the two-level interface from the other side. An operation must be specified in such a way that its inverse operation should be deduced from the in- and out-parameters. If some out-parameters are not important to the user of the type, these parameters may be hidden in the public interface.

Figure 1
A one-to-many relation from direct to inverse operations

| |
|---|
| **inverses**<br>   **for procedure** INSERT   (**in** Item;<br>                             **out** Report)<br>     **when** Report = AlreadyIn<br>       **inverse is** NULL<br>   **when** Report = Ok<br>     **inverse is** DELETE (Item, _ ) |

As two or more conceptually different private operations can be invoked through a common operation, e. g., the two conceptual INSERTs are merged in the same public INSERT operation, we provide a one-to-many relationship from private operations to inverse operations depending above all on out-parameters (Figure 1). Consequently, the inverse operation of a given operation cannot, (generally), be determined before the end of the execution of the direct operation.

*NULL inverse operation*
A particular inverse operation of interest is NULL. It is naturally associated as the inverse of any non modifying operation, e. g., CARD, (which returns the cardinal of a set), or IN, (which checks the membership). It is also associated to writing operations when it can be deduced from their out-parameters that they actually do not change the object state, as an INSERT which reports AlreadyIn, or an AND(false) on a boolean which was previously false. At last, it can even be associated to modifying operations when these modifications are irrelevant for the semantics of the ADT, e. g., a tree can be balanced for more efficient accesses if its organization is hidden (no ROOT or DEPTH operations, for instance).

The practical advantage in determining that the inverse of an operation is NULL, is that the direct and/or inverse operations need not be logged[1], (but the direct operation still has to be used for concurrency control).

### 2.3. Mapping public to private operations

We can also take advantage of this two-level interface for concurrency control by introducing more than one private operation for each public one, depending on the values of the in-parameters. The advantages are two-fold: conditional commutativity is simplified, and some operations might be recognized as unnecessary.

Let us consider the REAL type: It is possible to associate with the public MULTIPLY operation either the private MULTIPLY, DIVIDE, or SETTO operations, or the NULL operation; to the public ADD operation we associate either the private ADD, or SUB operations, or again the NULL operation (Figure 2).

*Simplified conditions*
Commutativity conditions are simplified. In this example, the amounts for the private ADD and SUB operations are always strictly positive values, a pre-condition often implicitly assumed.

The benefit is even larger if we consider the MULTIPLY operation which can be derived as a SETTO(0). First, this shortens the commutativity conditions between MULTIPLY and other operations since a cumbersome case is eliminated. Secondly, the determination of the inverse operation is also simplified! When multiplying by zero, the only way to get back to the previous state of the object is to restore its so-called before-image: there is no mathematical inverse. This will be managed directly by the SETTO operation and not as a special case of the MULTIPLY one.

Figure 2
A one-to-many relation from public to private operations

```
public
  procedure MULTIPLY (in Mult) is
    when Mult = 0 do
      private SETTO (0)
    when Mult = 1 do NULL
    when Abs(Mult) ≥ 1 do
      private MULTIPLY (Mult)
    else  do private DIVIDE (1 / Mult)
  procedure ADD (in Amount) is
    when Amount > 0 do
      private ADD (Amount)
    when Amount < 0 do
      private SUB (— Amount)
    else  do NULL
```

*NULL direct operation*
NULL deserves special attention because it is an operation which does exactly nothing. NULL has the property to commute with every operation. Therefore, it can be totally ignored, i. e., it does not need any checking with previous or following operations!

We dwell upon the fact that this is only true for operations which translate to NULL when considering only their in-parameters, i. e., direct NULL operations must be deducible *a priori*; it is incorrect to conclude that an operation is NULL in regard of its out-parameters, because the operation has had to access to the object. For instance, a SETTO operation is never NULL even if the old value of the object is equal to the assigned one; in that case, it can only be concluded that this SETTO was equivalent to IDENTITY. This means that its *inverse* operation is NULL.

---

[1] We assume that a log mechanism is used for recovery, which seems to be the best mechanism [1], or, at least, the most implemented one.

### 2.4. Advantages of the two-level interface

The two-level interface is useful to deal with recovery as well as with concurrency control. Its main advantage for concurrency control is to simplify commutativity conditions, and for recovery to reveal what is necessary for undoing the direct operation, which reacts again upon concurrency control (since finding out the out-parameters necessary to the inverse operations makes clear the commutativity conditions between these operations).

The two-level interface also reveals us the two kinds of NULL operations: inverse and direct ones. The inverse NULL operation implies that the operation needs not be recorded in the log in order to undo its corresponding direct operation. The direct NULL operation is more novel, its main advantage is to need no concurrency control at all.

As a last example of the effectiveness of the two-level interface, let us look at the BOOLEAN type: six public operations, AND, OR, XOR, NOT, SETTO, and READ, are translated to three private operations, NOT, SETTO, and READ, (plus NULL), because OR(false), AND(true), and XOR(false) do not change the value of the boolean, and OR(true), AND(false), and XOR(true) are translated respectively to SETTO(true), SETTO(false), and NOT. Note that if the operations were written in order to return the new value of the boolean, instead of translating to NULL, they would have been transformed into READ.

### 3. EXPLOITING COMMUTATIVITY

### 3.1. Some characteristics

Utilizing in- and out-parameters of operations increases concurrency by eliminating *pseudo*-conflicts. Therefore, we use the parameters in order to describe commutativity between operations. We describe commutativity between *private* operations: Firstly, only private operations are executed. Secondly, we have seen in section 2.2. that inverses (in fact, out-parameters) play a part in determining commutativity of operations.

### 3.1.1. Non-state-based commutativity

However, we refuse to consider the current state of an object in order to describe commutativity.

An advantage of using a non-state-based approach is that the representation of an object and the implementation of its operations do not interfere with commutativity conditions, as long as the parameters of the private operations remain the same. This is exactly the same point of view as shared by developers of ADTs since they can alter the implementation of an ADT without causing troubles to its users.

Allowing concurrent executions of operations with checkings is another resulting benefit which we exploit in our protocol.

Finally, the state of an object is (just) an abbreviation of the sequence of operations performed on it since an initial value. Some results may be obtained without looking at the state of the object; our protocol includes the possibility to deduce out-parameters without actually executing the operation.

### 3.1.2. Commutativity between direct private operations only

We consider that commutativity has to be defined only between direct private operations. We use the common definition of commutativity, i. e., two operations commute if (1) the state of the object and (2) their respective out-parameters are not dependent of the execution order[2]. Because private operations return all the out-parameters necessary to undo the direct operation, this ensures that the system not only schedules serializable histories, but also makes rejects possible without cascading aborts. Brössler & Freisleben [3] have to check commutativity between direct and inverse operations to achieve this important objective.

---

[2] Here, we do not consider the case of non-deterministic operations.

Serializable and cascading rollback-free schedules are also obtained by Turc [20] and Badrinath & Ramamritham [2]. Their respective criteria are incomparable but both include the mere case where a writing does not commute after a reading but is allowed because the inverse of the reading is NULL. Either reading or writing can be rejected without influencing the other. However, a precedence dependency is created and the commits of the operations must take place in that order if both commit, when two-phase locking is used.

### 3.1.3. At most one execution of an operation

On the one hand, we would like to exploit all the parameters, in-parameters and out-parameters, but on the other hand, we do not wish to execute first an operation on an "optimistic" basis, as is done in Argus [21] [12]. Using commutativity and desiring that an operation be executed only once implies that that operation is executed only when it commutes with all the other active operations. Since this guarantee has to be obtained before executing it, that means that its own out-parameters cannot be used!

The rationale for this constraint is the following: whenever an operation is rejected, all the following operations done on an "optimistic" way must be undone, redone and rechecked. If the operations are lengthy and/or numerous, the overhead would be unacceptable. To explain carefully this point, we study two "optimistic" protocols.

In the "very optimistic" protocol, every operation invoked on an object is immediately executed. Operations are classified either as active operations or blocked ones, i. e., belonging to an active transaction, or being the last operation of a blocked one. The classification of a new operation is decided after its execution by testing commutativity with all the other operations, active as well as blocked ones. It is active if it commutes with all, otherwise it is blocked. Blocked operations are waiting on active operations, (direct blocking), or on other blocked operations, ("transitive" blocking); thus, they are "ordered[3]" by the blocking relation. Conversely, active operations are not related to each other, and "precede" every blocked operation.

A problem of atomicity, with lengthy operations, is that execution and classification must be done atomically. However, this is not the Achilles heel of this protocol which is quite interesting as long as there is no rejects. Otherwise, two problems appear: an overhead which may be high, and, as a side-effect, starvation.

Suppose that an active operation has to be undone, then we must also undo all the operations which are directly or transitively blocked on that operation, (in the reverse order of invocation). Next, the blocked operations have to be redone, (in the previous chronological order), and checked another time, i. e., introducing and/or removing dependencies between operations. This rechecking is necessary because out-parameters have changed for direct dependencies, and perhaps for transitive ones too. Since this heavy task, the complexity of which is in $O(n^2)$ where n is the number of operations, takes place at each abort, a given transaction may be undone, redone, and rechecked very often, even infinitely often.

Starvation appears because out-parameters change and consequently new dependencies appear. Furthermore, since only some blocked operations are undone and redone, the first chronological order of execution of the operations is not preserved. Thus, an operation may be continuously redone and new dependencies established with now preceding operations. Starvation can be eliminated at the price of undoing and redoing all the blocked operations.

---

[3] The blocking (resp. waiting) relation *is not* a partial order.

A derived protocol can be proposed. This time, at most one blocked operation is executed in an "optimistic" way. When a new operation arrives, if there is no blocked operation, it is executed and checked, if there is already one blocked operation, the new one waits for it. We then have a set of pairwise commutative operations, a single blocked operation on some of the active ones, and a set of waiting operations, (on a FIFO queue), all waiting on the blocked one. The cost of one reject is constant: there is just one operation to undo, redo, and recheck. In our opinion, this second approach can possibly be used if rejects are not frequent or operations very short.

### 3.1.4. Deduced out-parameters

Executing only once an operation implies that out-parameters of this operation are missing for checking commutativity, thus the number of conflicts is increased. This drawback can be overcome to some extent by deducing out-parameters from the out-parameters of previous and still active operations.

The specifications of ADTs imply that when an operation has returned some values, the results of other operations are related to them. Such implications can be found in numerous conditional commutativity tables: for SET, if an INSERT(X) reports AlreadyIn, we know that a subsequent INSERT(X) has to report AlreadyIn too; for STACK, EMPTY reporting Yes implies that POP will report EmptyStack, (and conversely).

Unfortunately, there is still some serializable executions that are denied, e. g., in a SET, CARD will prohibit an incoming INSERT or DELETE, though an INSERT reporting AlreadyIn or a DELETE reporting NotFound commute with CARD. This limitation is due to the fact that sound deductions can be done only between operations which actually do not change the state of the object.

### 3.2. The protocol

Our protocol is not new at the transaction level; we use the well-known two-phase locking (2PL) protocol [5]. Furthermore, we choose the strict variation to avoid the cascading-rollback problem.

What is novel is the protocol adopted for executing an operation on an object. This protocol is divided into four steps: trying to (1) deduce out-parameters; then, if it is not possible, (2) executing an in-control, followed by (3) the execution of the operation itself, and at last (4) an out-control. These steps are illustrated in Figure 3 which describes the generic code associated to an operation.

As already stated, concurrency control and recovery take place at the private interface level, i. e., only private operations are involved.

The four steps of the execution of an operation must be atomic, independently however. The steps (1), (2), and (4) are critical sections managed by a monitor mechanism (Figure 7), while step (3) can be controlled by another mechanism in order to increase concurrency. For complex objects, such as b-trees, it is unthinkable to limit concurrency by executing the operations, i. e., step (3), in mutual exclusion.

Figure 3
Generic code for an operation

```
public procedure OP (in IN;  out OUT) is
    // select and call a private OP
    // translate private OUT to public OUT

private procedure OP (in IN;  out OUT) is
    // (1) deduce OUT
    if not Deduced
    then
        // (2) execute InControl
        // (3) execute private OP itself
        // (4) execute OutControl
```

We now detail the implementation of steps (1), (2), (4), and of an additional step for committing or rejecting the operations.

### 3.2.1. Data structures of the monitor

All the operations on an ADT are classified either as active or blocked. Between active operations, we distinguish already executed ones and in-execution ones. This distinction is necessary, because out-parameters are only available for executed operations. Thus, we have three sets of operations (Figure 4) which partitions the set of invoked and not yet rejected or committed operations.

The monitor also maintains two almost symmetrical relations: blocks and waiting-for. The "blocks" relation associates to any invoked operation the set of blocked operations which do not commute with it. The "waiting-for" relation associates to a given blocked operation the number of invoked operations which do not commute with it and precede it in the chronological order of invocation.

In the sequel, we omit a discussion of step (1) for the sake of genericity. Furthermore, it is worth merging step (1) with steps (2) and (4) in an implementation.

Figure 4
partition of invoked operations

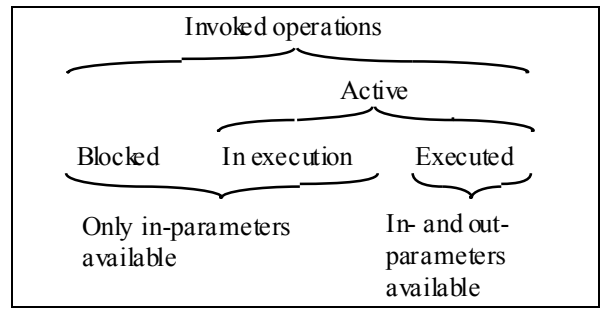

### 3.2.2. In-control

An operation is never executed if we are not sure of its commutativity with all the previous operations. Thus, when a new operation is invoked on an object, we have to check its commutativity with all the other operations, actives but blocked too.

For the new operation, solely the in-parameters are available, and the same for the blocked and in-execution operations. Therefore, conditional commutativity between the new operation and the union of blocked and in-execution operations is done on behalf of the in-parameters.

In Figure 5, we give the commutativity conditions using only in-parameters for a STACK ADT with four operations: PUSH, POP, EMPTY, CLEAR. Note that commutativity with only in-parameters is rather restricted.

Figure 5
In-commutativity for STACK ADT

```
in-commutativity
    commute PUSH (in x1)
        with PUSH (in x2) if x1 = x2;
    commute EMPTY (out report1)
        with EMPTY (out report2);
```

Conditional commutativity between the new operation and already executed ones can take into account the out-parameters of the executed operations. The conditions for the STACK example are given in Figure 6, where the first operation is the executed one. Note that all the conditions result in deduced out-parameters!

Figure 6
Out-commutativity for STACK ADT

> **out-commutativity**
>   **commute** POP (**out** x1;  **out** report1)
>   **with** POP (**out** x2;  **out** report2)
>     **if** report1 = EmptyStack
>     **then** report2 := EmptyStack;
>   **commute** POP (**out** x1;  **out** report1)
>   **with** EMPTY (**out** report2)
>     **if** report1 = EmptyStack
>     **then** report2 := true;
>   **commute** POP (**out** x1;  **out** report1)
>   **with** CLEAR (**out** report2)
>     **if** report1 = EmptyStack
>     **then** report2 := AlreadyEmpty;
>   **commute** EMPTY (**out** report1)
>   **with** POP (**out** x2;  **out** report2)
>     **if** report1 = true
>     **then** report2 := EmptyStack;
>   **commute** EMPTY (**out** report1)
>   **with** EMPTY (**out** report2)
>     **then** report2 := report1;
>   **commute** EMPTY (**out** report1)
>   **with** CLEAR (**out** report2)
>     **if** report1 = true
>     **then** report2 := AlreadyEmpty;
>   **commute** CLEAR (**out** report1)
>   **with** POP (**out** x2;  **out** report2)
>     **if** report1 = AlreadyEmpty
>     **then** report2 := EmptyStack;
>   **commute** CLEAR (**out** report1)
>   **with** EMPTY (**out** report2)
>     **if** report1 = AlreadyEmpty
>     **then** report2 := true;
>   **commute** CLEAR (**out** report1)
>   **with** CLEAR (**out** report2)
>     **if** report1 = AlreadyEmpty
>     **then** report2 := AlreadyEmpty;

The in- and out-commutativity descriptions correspond respectively to the CommuteWithIn and CommuteWithInOut boolean functions of the monitor code given in Figure 7.

### 3.2.3. Out-control

After executing the new operation, step (4) must take place. Its role is to remove some pseudo-conflicts which occurred during step (2) between blocked or in-execution operations and the new one. Since the new operation has been executed, its out-parameters are available. Then, out-commutativity is used between the new operation and the operations which are blocked on it.

### 3.2.4. Commit or reject

As we use a 2PL protocol, operations have to be committed at a later time, or rejected. Since a commit or reject implies to forget the operation and consequently to eliminate some dependencies between operations, this step has to be managed like steps (2) and (4).

In our basic protocol, this additional step is limited to removing dependencies and executing no longer blocked operations. Note that we use an extended signal primitive, able to unblock any given operation and not only the first one.

If this step is executed on behalf of a reject, then, for correctness reasons, it must be done after the execution of the inverse operation.

Figure 7
Monitor for the basic protocol

> **monitor** GeneralProtocol **is**
> **var**
>   Blocked:  **set of** Op;
>   InExecution:  **set of** Op;
>   Executed:  **set of** Op;
> *// The three categories of operations*
>   WaitingFor:  **map of** Op **to natural**;
> *// The number of operations waiting for the*
> *// commit or reject of an Op*
>   Blocks:  **map of** Op **to set of** Op;
> *// The set of operations blocked by a given*
> *// executed or in-execution Op*
>
> **entry** InControl (**in** NewOp: Op) **is**
> *// Called before execution of NewOp*

```
   loop for Op □ Blocked ≈ InExecution do
      if not CommuteWithIn(NewOp,Op)
      then
         WaitingFor(NewOp) += 1;
         Blocks(Op) ≈= {NewOp};
      end if;
   end loop;
   loop for Op □ Executed do
      if not CommuteWithInOut(NewOp,Op)
      then
         WaitingFor(NewOp) += 1;
         Blocks(Op) ≈= {NewOp};
      end if;
   end loop;
   if WaitingFor(NewOp) > 0
   then
      Blocked ≈= {NewOp};
      wait(NewOp);
      Blocked -= {NewOp};
   end if;
   InExecution ≈= {NewOp}.

entry OutControl (in NewOp: Op) is
// Called after execution of NewOp
   InExecution -= {NewOp};
   Executed ≈= {NewOp};
   loop for Op □ Blocks(NewOp) do
      if CommuteWithInOut(Op,NewOp)
      then
         WaitingFor(Op) -= 1;
         if WaitingFor(Op) = 0
         then
            signal(Op);
         end if;
         Blocks(NewOp) -= {Op};
      end if;
   end loop.

entry CommitOrReject (in NewOp: Op) is
// Called after execution of the
// inverse of NewOp, if it is a reject
   loop for Op □ Blocks(NewOp) do
      WaitingFor(Op) -= 1;
      if WaitingFor(Op) = 0
      then
         signal(Op);
      end if;
```

```
   end loop;
   Executed -= {NewOp};
   domain of Blocks -= {NewOp}.

init
   Blocked := Ø;
   InExecution := Ø;
   Executed := Ø;
   domain of WaitingFor := Ø;
   domain of Blocks := Ø;
end monitor.
```

## 4. CONCLUSION

In this paper, we have presented the general aspects of our framework for designing concurrent and recoverable ADTs. The correctness criteria remains serializability, and we use conditional commutativity.

The advantages of our scheme are: concurrency control and recovery are managed together; recovery is done through the use of inverse operations rather than before-images, which increases concurrency; commutativity is simplified by the use of the two-level interface; cascading rollback-free schedules are ensured; each operation is executed exactly once. This last advantage joined with the fact that conditional commutativity is not symmetrical leads to the major disadvantage: some couple of operations can be executed in one order but not in the reverse one.

In literature, a number of typical objects of interest are queues, stacks, counters, sets, files, directories, database relations, bank accounts, flight reservations, and so on. To manage efficiently all of these types of objects, a number of features have been added to this general framework. Some of them are transient data associated to an object in the monitor, commit-time operations, iterators [17].

The system is currently under implementation. We expect to learn from it which of these mechanisms are convenient,

i. e., "easy" to implement and efficient at run-time, the goal being always to allow better concurrency between transactions at the lowest cost.

The aspects related to nested transactions have been intentionally omitted in this paper because this proposition is independent of nesting.

**Acknowledgments**

We sincerely acknowledge Michèle Cart, Jean Ferrié, and Jean-François Pons for reading and discussing preliminary versions of this paper.